\documentclass[useAMS,usenatbib]{mn2e}

\usepackage[psamsfonts]{amssymb}
\usepackage[dvips]{graphicx}
\usepackage{amsmath,alltt}
\usepackage{multirow}
\usepackage{rotating}
\usepackage{hyperref}
\usepackage{pdflscape}
\usepackage{supertabular}

\title[Small-$N$ collisional dynamics IV]{Small-$N$ collisional dynamics IV:  Order in the realm of not-so-small-$N$}
\author[Leigh N. W. C., Geller A. M., Shara M. M., Baugher L., Hierro V., Ferreira D., Teperino E.]{Nathan W. C. Leigh$^{1,2,3}$, Aaron M. Geller$^{4,5}$,  
Michael M. Shara$^{1,6}$, 
\newauthor
Lukas Baugher$^{7}$, Vianny Hierro$^{7}$, De'Andre Ferreira$^{7}$, Elizabeth Teperino$^{7}$
\thanks{E-mail: nleigh@amnh.org (NWCL)}\\
$^{1}$Department of Astrophysics, American Museum of Natural History, Central Park West and 79th Street, New York, NY 10024 \\
$^{2}$Department of Physics and Astronomy, Stony Brook University, Stony Brook, NY 11794-3800, USA\\
$^{3}$Center for Computational Astrophysics, Flatiron Institute, 162 Fifth Avenue, New York, NY 10010, USA\\
$^{4}$Center for Interdisciplinary Exploration and Research in Astrophysics (CIERA) and Department of Physics and Astronomy, \\ 
Northwestern University, 2145 Sheridan Rd, Evanston, IL 60208, USA \\
$^{5}$Adler Planetarium, Dept.\ of Astronomy, 1300 S. Lake Shore Drive, Chicago, IL 60605, USA \\
$^{6}$Institute of Astronomy, University of Cambridge, Madingley Road, Cambridge CB3 0HA, UK \\
$^{7}$Student Research and Mentoring Program, Department of Education, American Museum of Natural History, \\
Central Park West and 79th Street, New York, NY 10024}
\begin{document}

\pagerange{\pageref{firstpage}--\pageref{lastpage}} \pubyear{2016}

\maketitle

\label{firstpage}

\begin{abstract}
In this paper, the fourth in the series, we continue our study of combinatorics in chaotic Newtonian dynamics.  We focus once again on the chaotic four-body problem in Newtonian gravity assuming finite-sized particles, and interactions that produce direct collisions between any two particles.  Our long-term goal is to predict the probability of a given collision event occurring over the course of an interaction, as a function of the numbers and properties of the particles.  In previous papers, we varied the number of interacting particles, as well as the distributions of particle radii and masses.  Here, we refine the methods developed in these preceding studies, and arrive at a final and robust methodology that can be used to study collisional dynamics in a variety of astrophysical contexts, ranging from stars in star clusters, galaxies in galaxy groups and clusters and even the collisional growth of planetesimals in protoplanetary disks.  We further present and refine the concept of a Collision Rate Diagram (CRD), the primary tool we use to quantify the relative rates for different collision scenarios to occur.  The agreement between our final theoretical model and the results of numerical scattering simulations is excellent.

\end{abstract}


\section{Introduction} \label{intro}

Particle-particle collisions during small-number fewbody interactions are the cause for several ubiquitous astrophysical phenomena.  These include, but are not limited to, blue straggler formation in globular and open star clusters due to stellar collisions \citep[e.g.][]{leonard89,fregeau04,leigh07,hypki16,hypki17}, the production of anomalously blue stars in galactic nuclei \citep[e.g.][]{shara74,davies98,bailey99,yu03,dale09,leigh16b}, the formation of intermediate-mass black holes via runaway stellar collisions that could also serve as the seeds for the formation of supermassive black holes \citep[e.g.][]{portegieszwart04,giersz15,stone17}, the collisional growth of protoplanetary disks \citep[e.g.][]{goldreich04,lithwick07}, the production of runaway stars from young star-forming regions \citep[e.g.][]{blaauw54,perets12,oh15,ryu17a,ryu17b,ryu17c}, the origins of elliptical galaxies \citep[e.g.][]{binney87,balland98,trinchieri03}, etc. 


Here, we continue our study of direct collisions between particles during chaotic few-body interactions.  We paraphrase the results of our previous works in this series here for completeness.  In Paper I \citep{leigh12}, we studied how the collision probability depends on the number of interacting particles.  We found a connection between the mean free path approximation and the binomial theorem. We showed that, for identical particles and a given total encounter energy and angular momentum, the collision probability scales roughly as $N^2$, where $N$ is the number of interacting particles.  The physical origin of this $N$-dependence comes from the binomial theorem; the number of ways of selecting any pair of particles from a set of $N$ identical particles is ${N \choose 2} = N(N-1)/2$.

In Paper II \citep{leigh15}, we found that, for (near-)identical mass particles, the collision probability is directly proportional to the collisional cross-section for the types of small-number interactions expected to occur in actual star clusters.  The dynamics of such gravitationally-bound systems of chaotically-interacting finite-sized particles are analogous to a system of pendulums; the particles oscillate semi-periodically about the system centre of mass.  Here, the cross-section for any two particles to collide directly is, to first order for particles with similar masses and large radii, proportional to the square of the sum of their radii.  By means of a combinatorics-based approach, it follows that the collision probability can be expressed analytically for any number of particles and any combination of particle radii.

In Paper III \citep{leigh16}, we derived analytic formulae for the time-scales for different collision scenarios to occur, and compared the results to numerical scattering simulations of binary-binary interactions.  We showed that the \textit{simulated} relative probabilities for the different collision scenarios are bounded by the corresponding analytic predictions, assuming either purely radial or purely tangential motions for the particles.   We further showed that, in the purely radial limit, our analytic time-scales provide good order-of-magnitude estimates for the mean time-scales for direct collisions to occur in our simulations.

In this paper, the fourth in the series, we study the probabilities for different collision scenarios to occur, while simultaneously varying the distribution of particle masses \textit{and} radii.  We first describe the framework underlying our model in the Newtonian limit, which is founded on a combinatorics-based backbone and is designed to calculate the time-scales or rates for direct collisions to occur during chaotic gravitational interactions involving finite-sized particles with different (but comparable) masses.  

In Section~\ref{method}, we apply the mean free path approximation to derive theoretical collision time-scales and \textit{relative} collision probabilities.  We further introduce the Collision Rate Diagram (CRD), which illustratively quantifies how well our derived collision time-scales are able to reproduce the simulated data.  We also describe the simulations used in this study to test our model.  In Section~\ref{results}, we present and compare the resulting simulated and theoretical collision probabilities and rates, using Collision Rate Diagrams as a guide to constrain to constrain the dominant physics needed to be reproduced via our analytic model.  The assumptions and limitations underlying our model are discussed along with their applicability to astrophysical systems in Section~\ref{discussion}.  Our key results are summarized in Section~\ref{summary}.
 
 \section{Method} \label{method}

In this section, we re-visit the concept of a Collision Rate Diagram (CRD), first presented in Paper III of this series. In this previous study, we derived different collision time-scales, and from these the various rates and probabilities for different collision scenarios to occur.  This will ultimately facilitate our ability to quantify in this paper, via the CRD, the effects of incorporating different assumptions in deriving these time-scales and rates.  We go on to present the numerical scattering experiments of binary-binary encounters involving finite-sized particles with different combinations of particle masses and radii.  We compare the results of these simulations to our analytic predictions in Section~\ref{results}, for different input assumptions to our model (e.g., setting the collisional cross-section equal to the geometric cross-section, setting it equal to the gravitationally-focused cross-section, with/without the assumption of time-averaged virial equilibrium, etc.), in order to identify the dominant physical processes deciding the relative collision rates and probabilities.  

Throughout this paper, we define a direct collision as occurring when the particle radii overlap directly, following the "sticky-star" approximation.  
 
\subsection{Model} \label{model}

In this section, we first re-visit the concept of a Collision Rate Diagram, before going on to present the numerical scattering experiments performed in this paper.  Later, these will be used, in conjunction with the CRDs presented in this section, to identify the dominant physical processes deciding the relative collision rates as a function of the the distribution of particle masses and radii, etc.

\subsubsection{Collision Rate Diagram} \label{CRD}

In this section, we re-introduce the concept of a Collision Rate Diagram (CRD), first presented in Paper III of this series.  This diagram provides an immediate and visual comparison between the predictions of our analytic derivations for the relative rates of collisions between different particle types (and their underlying assumptions; see Paper III) and the results of numerical scattering experiments.  This is because the area corresponding to a particular collision event is directly proportional to the probability of that outcome occurring.  Hence, as we will show, it provides a fast and efficient means of comparing theoretical predictions to simulated data.  By changing the assumptions underlying a given model, the parameter space indicating the rate dominance for each collision scenario will change.  Hence, in this way, the CRD is a potentially efficient tool for isolating the dominant physics deciding the relative rates for different collision scenarios to occur, and can be robustly applied to any astrophysical problem that touches upon the collisional regime of gravitational dynamics.

We begin by describing the original CRD from \citet{leigh16}, and repeat it here for completeness.  First, consider interactions involving three different types of particles, labeled A, B, and C.  In this case, we can use our derived time-scales to construct a Collision Rate Diagram, using a similar formalism as outlined in Paper III, and earlier in \citet{leigh11} and \citet{leigh13}.  A CRD is a diagram that illustrates the parameter space for which the rates of the different types of collisions (e.g., A+A, A+C, B+C, etc.) each dominate over all others.  In Paper III of this series, we considered only 3-dimensional CRDs, which have only a single quadrant, facilitated by writing the fraction of single stars as the sum of the fractions of binaries and triples, combined with the critical assumption of mass conservation for the entire system.

The procedure for producing such a CRD is as follows.  First, we note that, for three particle types, we can write the total number of particles $N$ involved in an interaction as:
\begin{equation}
\label{eqn:number}
N = N_{\rm A} + N_{\rm B} + N_{\rm C},
\end{equation}
where $N_{\rm A}$, $N_{\rm B}$ and $N_{\rm C}$ denote, respectively, the total number of particles of type A, B and C.  Then, the fraction of objects of a given particle type $i$ can be written:
\begin{equation}
\label{eqn:fraction1}
f_{\rm i} = \frac{N_{\rm i}}{N},
\end{equation}
and the sum of their total must of course satisfy the relation:
\begin{equation}
\label{eqn:fraction2}
1 = f_{\rm A} + f_{\rm B} + f_{\rm C}
\end{equation}

Now, to produce a CRD, every pair of collision rates (e.g., $\Gamma_{\rm A+A}$, $\Gamma_{\rm A+B}$, $\Gamma_{\rm A+C}$, etc.) should be equated, and the resulting relation plotted in $f_{\rm B}$-$f_{\rm C}$-space.  The region of parameter space in the $f_{\rm B}$-$f_{\rm C}$-plane for which each type of collision occurs at the highest rate can then be identified, and a corresponding boundary can be drawn in the CRD.  This produces a diagram that identifies the parameter space in the $f_{\rm B}$-$f_{\rm C}$-plane for which the rates of the different types of collisions (e.g., A+A, A+C, B+C, etc.) each dominate.  

\begin{figure}
\begin{center}                                                                                                                                                           
\includegraphics[width=\columnwidth]{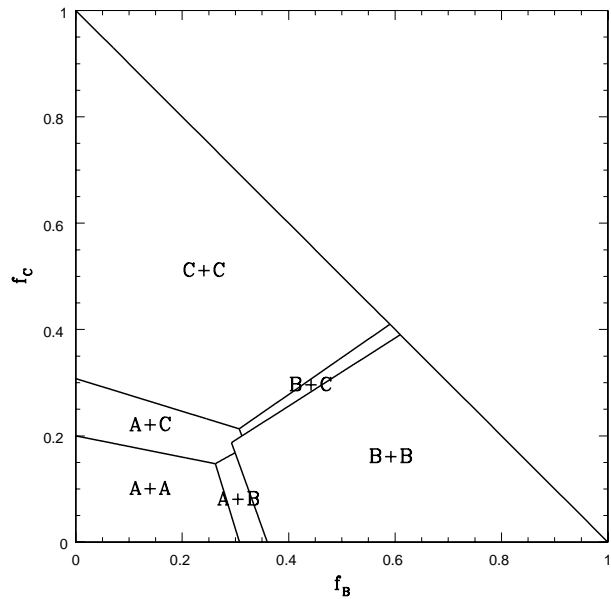}
\end{center}
\caption[A Collision Rate Diagram for three different types of particles]{The parameter space in the $f_{\rm B}$-$f_{\rm C}$-plane for which different types of collisions dominate.  To generate the CRD shown here, we assume three different types of particles with masses $m_{\rm A} =$ 1 M$_{\odot}$, $m_{\rm B} =$ 2 M$_{\odot}$ and $m_{\rm C} =$ 3 M$_{\odot}$, and radii $R_{\rm A} =$ 1 R$_{\odot}$, $R_{\rm B} =$ 2 R$_{\odot}$ and $R_{\rm C} =$ 3 R$_{\odot}$.  
\label{fig:fig1}}
\end{figure}

Figure~\ref{fig:fig1} shows an example of a 3-D Collision Rate Diagram.  To construct this figure, we assume for simplicity that the rate of collisions between particles of type $i$ and $j$ can be written:
\begin{equation}
\label{eqn:gammaij}
\Gamma_{\rm i+j} = f_{\rm i}f_{\rm j}N_{\rm i}n_{\rm j}{\sigma_{\rm i+j}}v_{\rm i+j},
\end{equation}
where we do not yet include gravitational-focusing in our estimate for the collisional cross-section $\sigma_{\rm i+j}$, and instead set it equal to the geometric cross-section for collision.  We emphasize that this is the rate for \textit{any} particle of type $i$ to collide with \textit{any} particle of type $j$.  This is in contrast to the rate for a \textit{given or specified} object of type $i$ to collide with \textit{any} particle of type $j$, which is smaller than the previous rate by a factor $N_{\rm i}$.

Equation~\ref{eqn:gammaij} can be modified to generate the simplest possible form for this CRD.\footnote{We note that Equation~\ref{eqn:gammaij} is itself not strictly correct, since the relative rates should include a combinatorial correction to avoid over-counting collisions between identical particles.  We will return to this later, and properly include this correction in our formulation.}  Specifically, we can assume the same particle number density $n = n_{\rm j}$ and relative velocity at infinity $v = v_{\rm i+j}$ for each type of collision (i.e., for all $i$ and $j$).  This simplifying assumption neglects the more complicated geometry considered in Paper III, but probes the simplest physical assumptions possible for the problem at hand (i.e., a suitable starting point for the method).  Hence, the simplified CRD shown in Figure~\ref{fig:fig1} quantities only the importance of the particle number and the geometric collisional cross-section in determining the relative collision rates.  

\begin{figure}
\begin{center}                                                                                                                                                           
\includegraphics[width=\columnwidth]{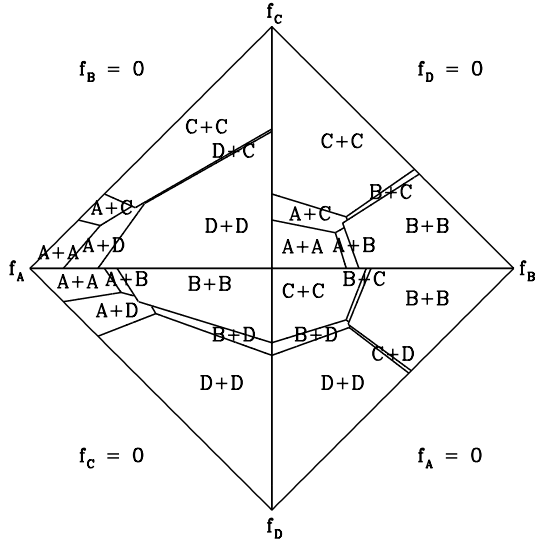}
\end{center}
\caption[A Collision Rate Diagram for four different types of particles]{Each quadrant shows the parameter space in the $f_{\rm i}$-$f_{\rm j}$-plane for which different types of collisions dominate, assuming one of the particle types is not present.  To generate the CRD shown here, we assume four different types of particles with masses $m_{\rm A} =$ 1 M$_{\odot}$, $m_{\rm B} =$ 2 M$_{\odot}$, $m_{\rm C} =$ 3 M$_{\odot}$ and $m_{\rm D} =$ 4 M$_{\odot}$, and radii $R_{\rm A} =$ 1 R$_{\odot}$, $R_{\rm B} =$ 2 R$_{\odot}$, $R_{\rm C} =$ 3 R$_{\odot}$ and $R_{\rm D} =$ 4 R$_{\odot}$.  
\label{fig:fig2}}
\end{figure}

Now, let us add a fourth particle type in to the mix, with label D.  Here, we can generate a 4-dimensional CRD, which has four quadrants and ultimately represents a 2-dimensional slice of the over-arching 4-dimensional parameter space.  In each of the four quadrants, we set the number of one particle type to be zero.  Then, each quadrant is effectively analogous to the CRD shown in Figure~\ref{fig:fig1}.  Combining all four quadrants allows for a more thorough comparison between theoretical predictions and the simulations for a larger subset of the total possible parameter space.  An example of a 4-D CRD is shown in Figure~\ref{fig:fig2}, adopting the same assumptions as in Figure~\ref{fig:fig1} in order to generate the simplest form of the CRD.

In the subsequent sections, we will compare the results of numerical scattering simulations to our analytic predictions.  This will be done using various forms of the CRD as our guide toward isolating the dominant physics deciding the relative collision rates or probabilities.  In this section, we have presented an over-simplified form of the CRD.  By introducing additional physics in to our model in the subsequent sections, specifically the gravitationally-focussed cross-section and a combinatorial correction, we will illustrate and quantify the effects of each of these physical components on our over-arching model.  

\subsection{Numerical scattering experiments} \label{exp}

As in Paper III of this series, we calculate the outcomes of a series of binary-binary (2+2) encounters using the \texttt{FEWBODY} numerical 
scattering code\footnote{The source code can be found at http://fewbody.sourceforge.net.}.  As discussed in more detail in \citet{fregeau04}, the code 
integrates the usual $N$-body equations in position-space in order to advance the system forward in time.  This is done using the 
eighth-order Runge-Kutta Prince-Dormand integration method with adaptive time-stepping and ninth-order error estimation.  

In this paper, we set $m_{\rm A} =$ 1 M$_{\odot}$, $m_{\rm B} =$ 2 M$_{\odot}$, $m_{\rm C} =$ 3 M$_{\odot}$ and $m_{\rm D} =$ 4 M$_{\odot}$, with $R_{\rm A} =$ 1 R$_{\odot}$, $R_{\rm B} =$ 2 R$_{\odot}$, $R_{\rm C} =$ 3 R$_{\odot}$ and $R_{\rm D} =$ 4 R$_{\odot}$.  We then consider different combinations of these four over-arching particle types.  

For these simulation sets, all particles are assumed to have finite radii (i.e., spherical) and we adopt the indicated combinations of masses and radii in Table~\ref{table:stats}.  In all simulations all binaries have $a_{\rm A} =$ $a_{\rm B} =$ 5 AU initially, and eccentricities $e_{\rm A} = e_{\rm B} =$ 0.  We set the impact parameter to zero and the initial relative velocity at infinity $v_{\rm rel}$ to 0.3$v_{\rm crit}$, where $v_{\rm crit}$ is the critical velocity.  It is defined as the relative velocity at infinity required for a total encounter energy of zero.\footnote{Note that this choice of relative velocity is typical for dense star clusters.}  As found in previous studies \citep[e.g.][]{leigh16}, such low relative velocities at infinity and small impact parameters maximize the probability of long-lived resonant interactions occurring, for which the assumption of ergodicity is upheld.  All angles defining the relative configurations of the binary orbital planes and phases are chosen randomly.  We perform 4 $\times$ 10$^4$ numerical scattering experiments for every combination of particle masses.  

As in previous papers in this series, all simulations are terminated at the instant the first collision occurs.  If no collisions occur, we use the same criteria as in \citet{fregeau04} to determine when an encounter is complete.  We refer the reader to \citet{fregeau04} and the previous papers in this series for the precise implementation of the stopping criteria used in this paper.  
As in previous studies, we adopt a tidal tolerance parameter 
$\delta =$ 10$^{-7}$ for all simulations.  The reader can refer to previous papers in this series for the full justification underlying this choice for $\delta$ (see also \citealt{geller15} and \citealt{leigh16}).

\section{Results} \label{results}

In this section, we present the results of our numerical scattering experiments and compare them to our theoretical predictions.  This is done by plotting the results of our numerical experiments in different manifestations of the Collision Rate Diagram, each time incorporating additional physics in to the model that generates the CRD.  These results are summarized below in Table~\ref{table:stats}, and illustrated in Figures~\ref{fig:fig3}, ~\ref{fig:fig4} and~\ref{fig:fig5}.  

\subsection{Confronting the analytic rates with simulated data} \label{compare}

\begin{figure}
\begin{center}                                                                                                                                                           
\includegraphics[width=\columnwidth]{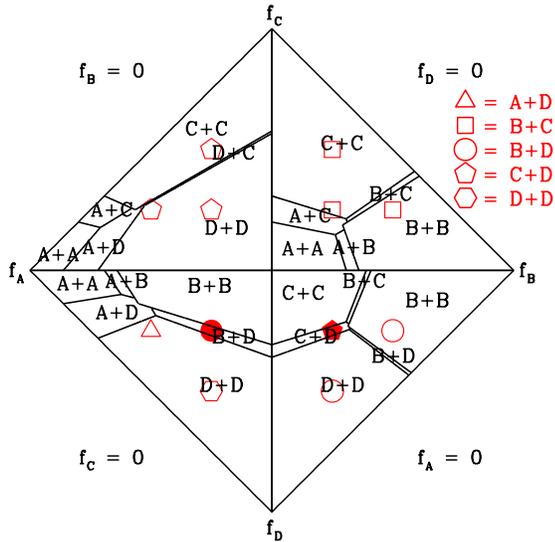}
\end{center}
\caption[A Collision Rate Diagram for four different types of particles, with the simulated data over-plotted]{The same as Figure~\ref{fig:fig2} but with the simulated data over-plotted.  
\label{fig:fig3}}
\end{figure}

In order to compare the results of the simulated data with the predictions of the CRD, we can include points in Figure~\ref{fig:fig3} for those combinations of $f_{\rm A}$, $f_{\rm B}$, $f_{\rm C}$ and $f_{\rm D}$ included in Table~\ref{table:stats} via our simulations.  The points are assigned shapes according to the outcome specified in each figure.  If the relative collision probabilities in the simulations agree exactly with the analytic predictions shown in the CRD, we plot the points as filled (i.e., the points fall in a region of the CRD corresponding to the dominant outcome found in our simulations).  If the points fall in an incorrect region of the CRD, we leave them unfilled.  

Figure~\ref{fig:fig3} clearly shows that the agreement between the simulations and the predictions of the CRD is poor without including additional physics, such as gravitational-focusing and a combinatorial correction.  If we include these additional effects in our model, can we improve upon the reported disagreement between theory and simulations?  This is directly addressed below.

\begin{figure}
\begin{center}                                                                                                                                                           
\includegraphics[width=\columnwidth]{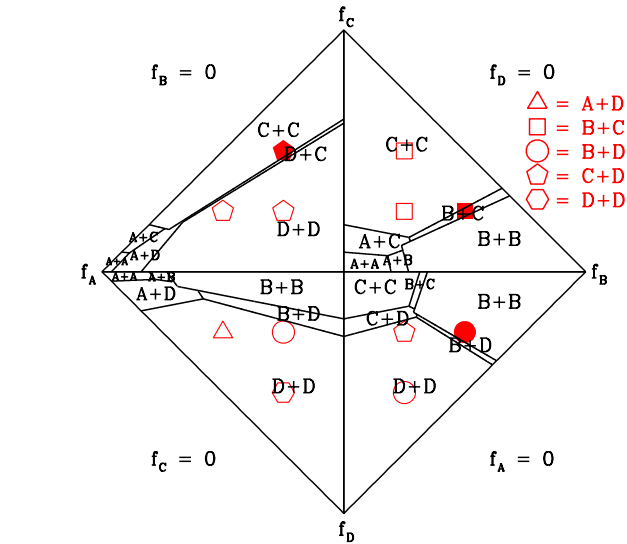}
\end{center}
\caption[A Collision Rate Diagram for four different types of particles, with the simulated data over-plotted]{The same as Figure~\ref{fig:fig3}, but adopting the gravitationally-focussed cross-section for collision instead of the geometric cross-section.
\label{fig:fig4}}
\end{figure}

To address this, we re-construct Figure~\ref{fig:fig3}, but adopting the gravitationally-focused cross-section for collision instead of the geometric cross-section.  This is motivated by the derived collision time-scales and rates presented in Paper III of this series.  As shown in Figure~\ref{fig:fig4}, the net effect of these alternations to the CRD is to increase the probability of collisions involving heavier particles.  The inclusion of this additional physics in our model only slightly improves the agreement between theory and simulations.
%

Finally, we include a combinatorial correction in our estimates for the relative collision rates in Equation~\ref{eqn:gammaij}.  Specifically, the number of ways of selecting two identical particles from a sample of $N$ identical particles is ${N \choose 2}$.  To account for this, we directly correct our assumed number fractions in Equation~\ref{eqn:gammaij} if $i = j$, and leave the rates unchanged otherwise.  This corrects for over-estimating the rate of collisions between identical particles in our simplified formulation presented in Section~\ref{CRD}.  

\begin{figure}
\begin{center}                                                                                                                                                           
\includegraphics[width=\columnwidth]{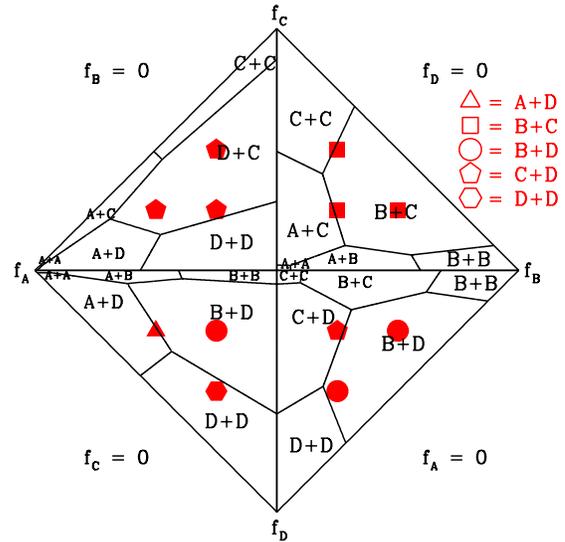}
\end{center}
\caption[A Collision Rate Diagram for four different types of particles, with the simulated data over-plotted]{The same as Figure~\ref{fig:fig4}, but adopting a combinatorial correction to the theoretically-derived collision rates, as discussed in the text.  
\label{fig:fig5}}
\end{figure}

As shown in Figure~\ref{fig:fig5}, this additional correction to our base model further improves the agreement between theory and simulations.  Now, all data points agree with our theoretical model.  It follows that our theoretical model successfully reproduces the simulated data for all of our simulations.

\subsection{The validity of the CRD for unexplored regions of parameter space} \label{unexplored}

We consider in total 12 combinations of particle types (A, B, C or D), but many more are also possible.  This choice was done for simplicity and to ensure that each simulated data point would appear only once in the CRD, in one of the four quadrants shown in Figure~\ref{fig:fig5}.  We do not include combinations of only two particle types, for example, since these simulated data points would be degenerate, and appear in more than one place in the CRD (i.e., two different quadrants).

With that said, we did perform a few additional scattering experiments assuming only two types of particles in a given four-body interaction.  Upon comparing the results of these simulations to our model and plotting them in the CRD, we consistently find excellent agreement between the theoretical predictions (i.e., CRD) and the simulations.  For example, we performed analogous simulations to those described in Section~\ref{exp} but assuming two particles of type B and two particles of type D.  The simulated data predict that B+D collisions should be dominant for $f_{\rm B} = f_{\rm D} = 0.5$.  A simple comparison to Figure~\ref{fig:fig5}, and specifically the lower right and left panels, reveals that the simulation outcomes are once again successfully described by the CRD.   Upon assuming two particles of type B and two particles of type C, the simulated data predict that B+C collisions should be dominant for $f_{\rm B} = f_{\rm C} = 0.5$.  Once again, a simple comparison to Figure~\ref{fig:fig5}, and specifically the upper and lower right panels, confirms that the simulation outcomes agree with the CRD.  We also point that, for any set of simulations assuming all identical particles, the resulting data points will always appear at the extremums of the various axes in Figure~\ref{fig:fig5}.  Consequently, by definition, these simulated data points will also always agree with the predictions of the CRD.  All of these examples further supports the validity the CRD developed and tested in this paper.


\clearpage
\begin{landscape} 
\begin{table}
\begin{center}
\centering
\begin{tabular}{|c|c|c|c|c|c|c|c|c|c|c|c|c|c|c|}
\hline
Particle  Combination    &   \multicolumn{10}{|c|}{Simulated Number}        \\
                  &      \multicolumn{10}{|c|}{of Collisions}                   &             \\
   ($N_{\rm A}$,$N_{\rm B}$,$N_{\rm C}$,$N_{\rm D}$)         &         $N_{\rm A+A}$   &     $N_{\rm B+B}$     &     $N_{\rm C+C}$     &     $N_{\rm D+D}$    &   $N_{\rm A+B}$   &     $N_{\rm A+C}$     &     $N_{\rm A+D}$     &      $N_{\rm B+C}$   &     $N_{\rm B+D}$     &     $N_{\rm C+D}$      &     \\             
\hline
 (2,0,1,1)  &   213  &  0  &   0  &   0  &   0   &   1165   &   2778    &   0   &   0   &  7042    \\
 (2,1,0,1)  &  408  &  0  &  0  &   0   &  1309  &  0   &   8071   &   0  &  5510  &   0   \\
 (2,1,1,0)  &  262  &   0   &   0  &   0   &   3305  &  5530   &   0   &   6150   &   0   &   0   \\
 (1,2,1,0)  &   0   &   2471   &   0   &   0  &    1600   &   1099   &    0    &    6711  &    0    &   0       \\
 (1,2,0,1)  &   0   &     638    &    0   &   0    &    1115   &    0    &    2954   &    0    &   8144   &    0       \\
 (0,2,1,1)  &  0  &  1071 &  0   &   0   &   0   &   0   &   0   &   2507   &   7628  &  5668  \\  
 (0,1,2,1)  &   0   &     0      &    2060    &    0     &    0     &    0      &     0     &     3134    &    3481    &    10839       \\
 (1,1,2,0)  &   0   &     0     &     6004     &     0       &     573     &      2970   &   0  &    7384   &  0   &   0       \\
 (1,0,2,1)  &  0  &   0    &   2943   &    0    &    0   &   1203   &   1224  &   0   &   0   &   11008  \\
 (0,1,1,2)  &  0  &  0  &  0  &  5907  &  0  &  0  &  0  &  1625  &  6177  &   6679  \\
 (1,0,1,2)  &  0  &   0  &   0  &   6535   &   0   &    519   &   1685   &    0    &   0   &  9151  \\   
 (1,1,0,2)  &  0   &   0  &  0  &  7469  &  58  &  0   &  1871  &   0   &   3965  &   0  \\
\end{tabular}  
\end{center}
\caption{The simulated numbers of each type of collision for different combinations of particle masses and radii.}
\label{table:stats}
\end{table}
\end{landscape}
\clearpage

\section{Discussion} \label{discussion}

In this paper, we present a method for directly comparing to simulated data the analytic rates for different collision scenarios to occur during chaotic gravitational interactions involving arbitrary numbers of finite-sized particles, with any distributions of particle masses and radii.  The method is flexible in the sense that the assumptions underlying the derivations of the analytic rates can be freely modified, and the subsequent impact on the agreement with the simulated data is easily quantified.  This facilitates a quick identification of the dominant physics deciding the relative rates of collisions, for the chosen initial conditions and particle properties.  To illustrate this, we compare our analytic predictions to the results of numerical scattering experiments of four-body interactions involving finite-sized particles.  Overall, the agreement between our analytic predictions and the simulations is excellent.  

This paper is meant as a more in-depth introduction to the CRD than given in Paper III of this series, while clearly articulating how to properly use it to construct a robust analytic framework for predicting collision rates, depending on the particle types and important physics (sticky-star approximation for collisions?  dissipation?  GW emission? etc.) for the problem to which the user wishes to apply their version of the CRD.  In other words, the validity of the CRD depends on the astrophysical environment of interest, since this in turn decides the types and properties of the particles (e.g., stars, comets, galaxies, etc.) in addition to the dominant physics affecting the collision rates (e.g., gravitational-focusing, combinatorics, various forms of dissipation such as from tides or GWs, etc.).


\subsection{Caveats} \label{caveats}

A few cautionary notes should be kept in mind when applying the Collision Rate Diagrams presented in this paper.  First, we expect our base model to be valid only for mass ratios $\lesssim$ 10 \citep{leigh16}.  Above this, we expect the most massive objects to sufficiently dominate the gravitational potential, such that the collision rate begins to enter the loss-cone regime (see \citet{merritt13} for more details) and/or the lowest mass particles are quickly ejected from the system without entering a resonant interaction state.  In general, a prolonged resonant state and the assumption of ergodicity being upheld are key requirements for our base model to be applicable \citep{leigh16}.  More work will be needed to better understand how to adapt our model to smoothly transition between these two regimes, where these assumptions start to break down.  Provided the assumption of ergodicity is upheld, we expect some variation of the collision rate estimates presented in Paper III to accurately capture the physics, in all but the most extreme cases (see below).

Finally, we note that more work is still needed to better understand the dependence of the collision probability on the impact parameter of the encounter, which was consistently set to equal zero throughout this paper.  An increase in the impact parameter should mostly act to increase the total angular momentum of the interaction (for a fixed relative velocity at infinity).  In future work, we intend to vary this parameter to better understand what adjustments to our base model might be needed to properly accommodate this additional free parameter.  The analytic model presented in Paper III of this series for the high-angular momentum regime, which assumes purely tangential motions relative to the system centre of mass, could provide the needed adjustments.

\subsection{Implications for dense stellar environments} \label{implications}

In this section, we describe the implications and applicability of our results to real astrophysical environments, with a focus on dense stellar clusters.

\subsubsection{Old globular clusters} \label{oldGCs}

Old globular clusters are the ideal environments for applying the methods presented in this paper.  There are three main reasons for this: 1) direct single-binary and binary-binary encounters (and possibly interactions involving triples as well) occur commonly in the dense cores of old GCs \citep[e.g.][]{leonard89,leigh11,leigh13}; 2) direct collisions between stars occur frequently during 3-body, 4-body, 5-body, etc. interactions in GCs \citep[e.g.][]{leonard89,leigh13,leigh15}; and 3) the range of stellar masses among such old stellar populations are typically $\sim 0.08 - 0.8$ M$_{\odot}$, which limits the mass ratios of the fewbody interactions to q $\gtrsim$ 0.1 and consequently ensures that the assumption of ergodicity (which is needed to apply our model and construct a CRD) should be approximately upheld for most chaotic fewbody interactions in old GCs.

To properly apply the CRD and methods presented in this paper to old GCs, a mass-radius relation for main-sequence stars should be adopted.  Typically, for such an old stellar population, the relation $R/R_{\odot} = (m/M_{\odot})^{0.75}$ is used, where $m$ is the main-sequence mass and $R$ is the corresponding stellar radius \citep{hansen04}.  This ensures that the particle masses and radii are chosen appropriately when constructing a CRD designed for chaotic fewbody interactions in old GCs.  Such a change would likely have the affect of reducing the importance of the particle radius in the CRD, such that the difference in area between the largest and smallest zones (e.g., D+D and A+A) would be reduced.

\subsubsection{Young massive clusters} \label{youngMCs}

Young massive clusters could also be well suited to applying the methods presented in this paper.  For example, \citet{portegieszwart04} showed using $N$-body simulations that runaway collisions of massive stars can occur in the dense cores of primordial GCs, possibly forming a supra-massive star or even an intermediate-mass black hole.  We emphasize that, since we are only interested in the \textit{relative} collision rates, the model is scale-free for our purposes if we fix the ratio of particle mass to particle radius such that it remains unity (as assumed in Figure~\ref{fig:fig5}, for example).  By extension, the CRD shown in Figure~\ref{fig:fig5} is equally well-suited to larger, more massive particles, provided the masses are all directly proportional to the masses assumed in Figure~\ref{fig:fig5}.  For example, if we replace the particles with masses of 1, 2, 3 and 4 M$_{\odot}$ with more massive particles with masses of 10, 20, 30 and 40 M$_{\odot}$, then Figure~\ref{fig:fig5} still applies provided the stellar radii are, respectively, 10, 20, 30 and 40 R$_{\odot}$.  Thus, our method is just as applicable to treating chaotic interactions involving more massive stars, provided the minimum mass ratio in each interaction satisfies $q \gtrsim 0.1$, as previously discussed.

\section{Summary} \label{summary}

In this paper, the fourth in the series, we push forward in our study of chaotic Newtonian gravity involving small numbers of finite-sized particles.  Our focus remains direct collisions between pairs of particles in the "sticky-star" approximation.  Our over-arching goal in this series of papers is to develop a method to calculate the probability of any two particles colliding during a chaotic (bound) resonant gravitational interaction involving any number $N$ of particles with any combination of particle masses and radii, as well as to directly compare its predictions to simulated data.  

In our previous papers, we showed that (1) the probability of a collision occurring during interactions involving identical particles is approximately proportional to $N^2$, which comes from combinatorics and, specifically, the number of ways of selecting any two particles from a larger set of $N$ identical particles, or ${N \choose 2}$ \citep{leigh12}; (2) for strongly bound gravitational encounters (i.e. with $E \ll$ 0, where $E$ is the total encounter energy, and having small impact parameters) involving small numbers of particles, the collision probability is directly proportional to the collisional cross-section.  For identical particle masses and large particle radii, the collisional cross-section is roughly equal to the sum of the cross-sectional areas of the colliding particles \citep{leigh15}; and (3) for different particle masses but identical particle radii, the mean free path approximation can be used in conjunction with (or without) the assumptions of time-averaged virial equilibrium and energy equipartition to derive estimates for the relative collision rates that agree with the simulated data at the order-of-magnitude level \citep{leigh17}.  

In this paper, we continue our study by considering interactions involving particles with \textit{both} different masses and radii.  As in Paper III, we consider the four-body problem in this paper.  This is because, for $N =$ 4, we can run more simulations since we minimize the computational expense due to the small number of particles.  This contributes to a significant increase in the statistical significance of the analysis.  Using our previous results from Papers I, II and III, we derive Collision Rate Diagrams for our case of interest.   This is done first assuming that only the geometric cross-sections for collision affect the relative collision rates, and then again assuming gravitational focusing and including a combinatorial correction.  For these cases, we analyze the \textit{relative} collision probabilities as predicted by our analytic formulae and compare them to the results of numerical scattering simulations performed with the \texttt{FEWBODY} code \citep{fregeau04}.  

By calculating different manifestations of the CRD, our method illustrates the following key result.  While the geometric cross-section captures most of the relevant physics for comparable particle masses, invoking the additional assumptions of a gravitationally-focused cross-section along with a combinatorial correction yield better agreement with the simulations.  With these additional assumptions, our analytic estimates reproduce the simulated data for \textit{all} combinations of particle masses and radii considered here.  

Our method (or variations thereof) is suitable to direct stellar collisions in dense star clusters, the collisional growth of planetesimals in protoplanetary disks, the growth of super-massive black holes via runaway stellar collisions, the formation of giant elliptical galaxies in the galaxy clusters and groups, and even the growth of large-scale structure in the Universe.

\section*{Acknowledgments}

The authors thank an anonymous reviewer for comments and suggestions that strengthened our manuscript.  N.~W.~C.~L. gratefully acknowledges support from the American Museum of Natural History and the Richard Guilder Graduate School, specifically the Kalbfleisch Fellowship Program, as well as support from a National Science Foundation Award No. AST 11-09395.  


\bsp

\label{lastpage}

\end{document}